\documentclass[notitlepage,twocolumn,showpacs,preprintnumbers,nofootinbib,superscriptaddress]{revtex4-1}
\usepackage{amssymb,amsmath,placeins,epsfig,array}
\usepackage[usenames,dvipsnames]{color}
\usepackage[normalem]{ulem}
\usepackage[breaklinks,colorlinks,urlcolor=blue,citecolor=blue,linkcolor=blue]{hyperref}

\usepackage[usenames,dvipsnames]{xcolor}

\newcommand{\ie}{{\textit{i.e.~}}}
\newcommand{\eg}{{\textit{e.g.~}}}

\usepackage{fancyhdr}

\begin{document}
\title{Electromagnetic Signatures of Dark Photon Superradiance}

\author{Andrea Caputo }
\thanks{These two authors contributed equally.}
\affiliation{School of Physics and Astronomy, Tel-Aviv University, Tel-Aviv 69978, Israel}
\affiliation{Department of Particle Physics and Astrophysics,
Weizmann Institute of Science, Rehovot 7610001,Israel}
\affiliation{Max-Planck-Institut f\"ur Physik (Werner-Heisenberg-Institut), F\"ohringer Ring 6, 80805 M\"unchen, Germany}
\author{Samuel J. Witte}
\thanks{These two authors contributed equally.}
\affiliation{Gravitation Astroparticle Physics Amsterdam (GRAPPA), Institute for Theoretical Physics Amsterdam and Delta Institute for Theoretical Physics, University of Amsterdam, Science Park 904, 1098 XH Amsterdam, The Netherlands}
\author{ Diego Blas}
\affiliation{ Theoretical Particle Physics and Cosmology Group, Department of Physics,
King's College London, Strand, London WC2R 2LS, UK}
\author{Paolo Pani  }
\affiliation{Dipartimento di Fisica, ``Sapienza'' Universit\`a di Roma \& Sezione INFN Roma1, Piazzale Aldo Moro 5, 00185, Roma, Italy}

\preprint{KCL-2021-06}

\begin{abstract} 
Black hole superradiance is a powerful tool in the search for ultra-light bosons. Constraints on 
the existence of such particles have been derived from the observation of highly spinning black holes, absence of 
continuous gravitational-wave signals, and of the associated stochastic background. However, these constraints are 
only strictly speaking valid in the limit where the boson's interactions can be neglected. In this work we 
investigate the extent to which the superradiant growth of an ultra-light dark photon can be quenched via scattering 
processes with ambient electrons. For dark photon masses $m_{\gamma^\prime} \gtrsim 10^{-17}\,{\rm eV}$, and for 
reasonable values of the ambient electron number density, we find superradiance can be quenched prior to  extracting a 
significant fraction of the black-hole spin. For sufficiently large $m_{\gamma^\prime}$ and small electron number 
densities, the in-medium suppression of the kinetic mixing can be efficiently removed, and quenching occurs for mixings 
$\chi_0 \gtrsim \mathcal{O}(10^{-8})$; at low masses, however, in-medium effects strongly inhibit otherwise efficient 
scattering processes from dissipating energy.  Intriguingly, this quenching leads to a time- and energy-oscillating 
electromagnetic  signature, with luminosities potentially extending up to $\sim 10^{56}\,{\rm erg / s}$, suggesting that 
such events should be detectable with existing telescopes. As a byproduct we also show that superradiance cannot be used 
to constrain a small mass for the Standard Model photon. 
\end{abstract}

\maketitle

\section{Introduction}

Black hole~(BH) superradiance is the process by which low-energy bosons can extract the rotational energy of a 
spinning BH~\cite{Press:1971wr,zeldovich1,zeldovich2,Press:1972zz,Teukolsky:1974yv,Zeldovich:1982zz} (see 
Ref.~\cite{Brito:2015oca} for an overview). 
This process is at play for modes with frequency $\omega < m \Omega$, where $\Omega$ is the angular velocity of the 
BH and $m$ the azimuthal quantum number of the mode. If this radiation is confined near the BH, the 
energy extraction may happen at an exponential rate, leading to a so-called `BH bomb' instability~\cite{Press:1972zz}.  
It has long been 
appreciated that the very mass of a particle can serve as such a confining 
mechanism~\cite{Brito:2015oca,Damour:1976kh,1977RSPSA.352..381D,Cardoso:2004nk}, as Kerr BHs have quasi-bound 
state orbits with support within and just outside the ergoregion for particles with Compton wavelength roughly comparable to the BH horizon.

This phenomenon has garnered much attention over the last decade as the search for exotic light particles, such as 
axions and dark photons, has intensified. Most studies thus far have focused on understanding the growth and 
evolution of a BH-boson condensate forming through the superradiant instability, under the simplifying 
assumption that the boson field is non-interacting. In this case, and if the boson mass $m_b \lesssim M^{-1}$ (being $M$ 
the mass of the BH; we use 
$G=c=\hbar=\kappa_B=1$ units hereafter), one expects the superradiant boson cloud to be able to extract up to $\approx 10\%$ 
of the angular momentum of a highly spinning BH over extremely short timescales~\cite{Brito:2014wla,Witek:2018dmd,East:2017ovw}. The 
presence of gaps in the BH mass-spin ``Regge'' plane could then serve as an indirect observation of the 
existence of such bosons~\cite{Arvanitaki:2009fg,Arvanitaki:2010sy}; equivalently, the observation of highly spinning BHs can be used to 
constrain the existence of exotic particles~\cite{Palomba:2019vxe,Zhu:2020tht,Brito:2017wnc,Brito:2017zvb,Tsukada:2018mbp,Tsukada:2020lgt,Brito:2020lup,Ghosh:2021zuf}. In addition, if the boson cloud remains bound to the BH after the 
superradiant condition has saturated (\ie when the BH spin has been reduced to a point where $\omega \simeq m 
\Omega$), its (spinning) dipolar structure produces nearly continuous quadrupolar gravitational waves at a frequency 
set by the boson mass. Negative searches for such continuous signals~\cite{Palomba:2019vxe,Zhu:2020tht} and for the stochastic 
background produced by unresolved sources~\cite{Brito:2017wnc,Brito:2017zvb,Tsukada:2018mbp,Tsukada:2020lgt} in LIGO and 
Virgo have set some further constraints on the mass of ultra-light 
bosons in a narrow range around $m_b\sim 10^{-13}\,{\rm eV}$~\cite{Brito:2020lup}. Future 
space-based interferometers such as LISA will probe much smaller masses~\cite{Brito:2017wnc,Brito:2017zvb} and could also detect other effects, such as the 
gravitational and environment effects of the bosonic cloud on the evolution of extreme mass-ratio inspirals~\cite{Ferreira:2017pth,Hannuksela:2018izj,Zhang:2019eid,Baumann:2019ztm}.
Overall, current and future (both electromagnetic and gravitational-wave) probes of superradiance can explore the ultra-light boson frontier roughly in 
the range $m_b\in(10^{-21},10^{-11})\,{\rm eV}$, see \cite{Brito:2015oca} for a summary of the latest constraints and for their 
dependence on the particle's spin.

The above description of the superradiant instability is strictly speaking valid only when the boson field is 
non-interacting. However, superradiance generates enormous number densities, and thus interactions that would 
conventionally be considered as weak may be sufficient to destroy or quench the evolution of the superradiant cloud. The 
investigation of quenching mechanisms, and complementary signatures that may arise even when the BH spin is not 
significantly altered, has received attention in recent 
years~\cite{Arvanitaki:2009fg,Arvanitaki:2010sy,Fukuda:2019ewf,Ikeda:2019fvj,Cardoso:2020nst,Blas:2020kaa,Blas:2020nbs,
Baryakhtar:2020gao}. Given the stringent nature of the superradiance constraints, and the enormous energy 
densities typically involved in the problem, understanding when and how various models undergo quenching is of utmost 
importance.

We focus here on the role of ultralight dark photon interactions during the superradiant growth, identifying parameters 
and model-dependent features for which quenching occurs, and illustrating that electromagnetic signatures may arise 
when quenching is important. Models featuring the existence of a dark photon are ubiquitous in extensions of the 
Standard Model~(SM) of particle physics~\cite{Holdom:1985ag,Jaeckel:2010ni,Essig:2013lka,Fabbrichesi:2020wbt}. 
Frequently referred to as the ``vector portal'', these models present a very generic possibility -- together 
with the Higgs and the neutrino portals -- to broadly characterize connections between the SM and dark 
sectors~\cite{PhysRevD.80.095024}. Over the last several years, the vector portal has become one of the most important 
paradigms in particle physics models, and is routinely invoked to explain existing experimental anomalies in high-energy 
physics and cosmology, see 
\eg \cite{Cline:2014dwa,ko2014higgs,Escudero:2017yia,Pospelov:2018kdh,Mohlabeng:2019vrz,Hooper:2019xss,Caputo:2020avy,Alonso-Alvarez:2020cdv,Bloch:2020uzh}. It is then relevant not to leave any stone unturned, and explore the diverse 
phenomenological consequences arising from BH superradiance in such models. 

We find that despite an initial in-medium suppression~\cite{Mirizzi:2009iz, Dubovsky:2015cca} of the interaction strength (especially relevant at low dark-photon masses), dark photons which kinetically mix 
with the SM photon will quench if their mass $m_{\gamma^\prime} \gtrsim 10^{-16} \, {\rm eV}$ and  vacuum mixing $\chi_0 
\gtrsim 10^{-7}$ (quenching may still occur for smaller mixings at larger masses, and larger mixings at masses down to 
$\sim 10^{-17}\,{\rm eV}$, however the details depend on the local electron number density which can span many orders 
of magntiude). 
Furthermore, we argue that the 
superradiant growth of such particles may give rise to a unique, time-dependent, electromagnetic signature that may be 
detectable using current telescopes. 

The outline of this paper is as follows. In Sec.~\ref{sec:VS} we present an overview of vector superradiance. We then 
present the dark photon model in Sec.~\ref{sec:DP}. In Sec.~\ref{sec:dps} we discuss the relevant scattering processes 
that are responsible for inhibiting the growth of the cloud, and estimate the luminosity from the semi-Compton 
scattering off, and synchrotron emission of, the ambient electrons. We then discuss additional model-dependent 
quenching 
mechanisms that do not rely on the existence of the kinetic mixing. Section~\ref{sec:sm} includes a brief discussion 
on implications for the SM photon, with particular reference to implications for superradiant bounds that have been 
derived on the potential existence of a bare photon mass. We conclude in Sec.~\ref{sec:con}.

\section{Vector superradiant instability}\label{sec:VS}
We consider the dynamics of a light vector boson $A^\prime$ with mass $m_{\gamma^\prime}$ on the background of a rotating BH with mass $M$ and dimensionless spin $\tilde{a} \equiv J/M^2$, where $J$ is the angular momentum of the BH.  Assuming that the vector field is non-interacting and neglecting backreaction on the metric, its evolution is governed by the Proca equation, $\nabla_{\sigma}F^{\prime \, \sigma \nu} - m_{\gamma^\prime}^2A^{\prime \, \nu} = 0$, on the Kerr spacetime. 
Originally, the superradiant instability in this case was studied in the Fourier domain and in the slow-rotation approximation~\cite{Pani:2013pma} up to ${\cal O}(\tilde a^2)$ by expanding the field in a basis of vector spherical harmonics with indices $(l,m)$, yielding a system of ODEs in which modes with different $l$ number and opposite parity are coupled to each other~\cite{Pani:2012vp,Pani:2012bp}. 
More recently, the eigenvalue problem was solved for arbitrary values of $\tilde a$ both analytically in the Newtonian approximation ($\alpha\equiv m_{\gamma^\prime} M\ll1$)~\cite{Baryakhtar:2017ngi,Baumann:2019eav}, and numerically in the generic case, either by solving a system of PDEs~\cite{Cardoso:2018tly,Baumann:2019eav} or by using the recently discovered separability~\cite{Frolov:2018ezx} of the Proca equations in the Kerr metric~\cite{Frolov:2018ezx,Dolan:2018dqv}.
The Newtonian approximation is typically sufficiently accurate to capture the order of magnitude of the instability time scale when $\alpha\lesssim0.1$. For a mode with $\omega = \omega_R + i \omega_I$, the real part and imaginary parts of the frequency are approximately given by~\cite{Pani:2012vp,Pani:2012bp,Baryakhtar:2017ngi}
\begin{eqnarray}
    \omega_R^2 \simeq m_{\gamma^\prime}^2\Big(1 - \Big(\frac{\alpha}{l + n + S +1}\Big)^2\Big),\\
    \omega_I = \frac{2\gamma_{Sl}}{M}r_+\Big(m\Omega-\omega_R\Big)\alpha^{4l + 5 + 2S},
\end{eqnarray}
where $r_+=M(1+\sqrt{1-\tilde a^2})$ is the horizon radius, $\Omega=\tilde{a}/(2r_+)$, $l=1,2,..$ is the total angular momentum number of the mode, $m$ is the azimuthal number (such that $|m|\leq l$), $n=0,1,2,...$ is the overtone number, $S=-1,0,1$ the polarization, and  $\gamma_{Sl}$ a numerical coefficient~\cite{Brito:2015oca}. The fastest growing mode corresponds to $S = -1, l =1$ (yielding $\gamma_{-11}=4$) and the superradiant timescale is approximately given by~\cite{Pani:2012vp,Pani:2012bp}
\begin{equation}\label{eq:timescale_super}
    \tau_{s} \simeq \frac{M \alpha^{-7}}{\tilde{a}\gamma_{-11}} \simeq \frac{10^2}{\tilde{a}} \Big(\frac{M}{10 M_{\odot}}\Big) {\rm s}\,,
\end{equation}
where in the last step we assumed a fiducial value $\alpha\sim 0.1$.
The typical radius of the superradiant cloud is roughly given by $r_{\rm cloud} \simeq M/\alpha^2$, and ranges from 
$r_{\rm cloud}\sim 100M$ for the fiducial value above down to $r_{\rm cloud}\sim 10M$ for $\alpha\sim 0.3$. 

If efficient, \ie if $\tau_s$ is small relative to the rate at which the BH's angular momentum grows (e.g. through 
accretion), the superradiant cloud will grow until it extracts sufficient amount of the angular momentum and saturates 
the superradiant condition; in the case of a nearly extremal BH, this occurs after the cloud has extracted up to 
$\approx10\%$ of the BH total energy~\cite{Brito:2014wla,Witek:2018dmd,East:2017ovw}. Given a cloud with mass $M_{\rm 
cloud}$, the angular momentum loss is $\Delta J= m M_{\rm cloud}/\omega_R$~\cite{Brito:2015oca}. Thus, assuming such 
light degrees of freedom exist, one should not observe highly spinning BHs with $M \sim (2 m_{\gamma^\prime})^{-1}$; or 
equivalently, since no such feature has been observed so far, the existence of highly spinning BHs can be used to 
constrain these light vector bosons. The BH accretion rate, controlling the characteristic time scale over which one can 
adopt a constant BH mass, can be conservatively estimated to be given by a fraction of the Eddington accretion 
timescale,
\begin{equation}
	\tau_{\rm Edd} = \frac{M}{\dot{M}} \sim \frac{\epsilon \sigma_T}{4\pi m_p} \sim 1.4 \, \epsilon \times 10^{15} \, {\rm s} \, ,
\end{equation}
where $\sigma_T$ is the Thomson cross section, $m_p$ the proton mass, and $\epsilon$ an efficiency factor. The time scale above with $\epsilon={\cal O}(1)$ can be considered as a conservative lower bound, since accretion can be much less efficient. We notice that the Eddington timescale is much larger than the superradiant timescale of interest, Eq.~\ref{eq:timescale_super}. Therefore we can safely neglect accretion in the rest of the work.

The superradiant growth can be impeded, or quenched, should processes exist that deplete the abundance of the ultra-light bosons at a rate faster than that of superradiant growth, or should the mass, and thus the bound states, of the boson be significantly modified. It is important to emphasize that the interaction responsible for the quenching does not need to be considered by conventional standards to be {\emph{strong}} -- the very large energy densities achieved during the superradiant growth can often compensate for highly suppressed scattering processes.

\section{The Dark Photon in a plasma}\label{sec:DP}

Dark photons interact with the SM through a kinetic mixing with the SM photon~\cite{Holdom:1985ag}. 
Working in a basis in which the quadratic action is diagonal, the Lagrangian reads
\begin{align}
\mathcal{L} =& -\frac{1}{4}F_{\mu\nu}F^{\mu\nu} - \frac{1}{4}F'_{\mu\nu}F'^{\mu\nu} + \\ & m^2_{\gamma'} A'_{\mu}A'^{\mu} - e J^{\mu}\Big(A_{\mu} + \sin\chi_0 A'_{\mu}\Big) \, , \nonumber
\end{align}
where $F_{\mu\nu}$ and $F^\prime_{\mu\nu}$ are the field strength tensors of the photon and dark photon, respectively, 
$\sin\chi_0$ is the bare kinetic mixing (assumed to be always much smaller than unity, $\sin\chi_0 \ll 1$), 
$m_{\gamma'}$ 
is the 
dark photon mass, and $J^{\mu}$ is a SM electric current. In this basis we therefore have a direct coupling between 
the hidden photon and the SM electric current. Notice that here we indicate by $A'_{\mu}$ the dark interaction eigenstate. The propagating dark state, found after the diagonalization of the mass matrix, will be instead named $A_{\rm dp}$.

For large dark photon masses, one can effectively decouple the Proca 
field solution from the intrinsic dynamics of the electrons and ions (that is to say, the motion of the electrons may be 
driven by the presence of the dark photon, but the dark photon is not affected by the presence of the plasma); in the 
small mass limit, however, the motion of these particles can induce in-medium effects~\cite{Mirizzi:2009iz, Dubovsky:2015cca} which cause \eg the dark photon 
and photon to fully decouple in the $m_{\gamma^\prime} \rightarrow 0$ limit~\cite{An:2013yua}. Thus, in this case it is necessary to jointly solve for the dynamics of the entire system\footnote{Recently the linearized dynamics of the SM 
electromagnetic field in a cold plasma and in curved spacetime has been studied in detail~\cite{Cannizzaro:2020uap}. 
Although the dynamics is quite rich, the final result in terms of quasi-bound states is in qualitative agreement with  what predicted by the dispersion relation of plane waves. In the following we shall therefore use the latter, much 
simpler approach, postponing a complete dynamical analysis of the system to future work.}.

The motion of the electrons (and ions, which we neglect in what follows due to the fact that their velocity is much lower than that of the electrons) in the presence of an electromagnetic field, including the effect of electron-ion collisions and assuming a cold plasma (\ie with temperature $T \ll m_e$), is given by 
 \begin{eqnarray}\label{eq:vlasov}
\frac{\partial \vec{p}_e}{\partial t}+ ( \vec{v}_e \cdot \nabla) \vec{p}_e = - e(\vec{E}+ \sin\chi_0 \vec{E}' +  \\ + \vec{v}_e \times (\vec{B} + \sin\chi_0 \vec{B}')) -  \nu(  \vec{p}_e -  \vec{p}_i) \, , \nonumber
\label{eq:electrons}
\end{eqnarray}
where $\vec{v}_e$ and $\vec{p}_e$ are the velocity and momentum of the electrons, $\vec{p}_i$ the momentum of the ions (which in the rest frame can be expressed as $-\vec{p}_e$), and $\nu$ is the electron-ion collision frequency. This equation must be solved simultaneously with the electromagnetic and Proca equations, given by
\begin{eqnarray}
	\Box A^\mu & = & e J^\mu \,,\\
	\Box A^{\prime \, \mu} + m_{\gamma^\prime}^2 A^{\prime \, \mu} & = & e \sin\chi_0 \, J^\mu \, .
	\label{eq:Proca_equations}
\end{eqnarray}
In full generality this problem is intractable, and thus in what follows we identify various limiting regimes of interest and use these solutions to determine how the dark photon field will evolve.

\subsection{Collisional Regime}
Let us first consider the effect of electron-ion collisions (note that the propagation of the ultralight dark photon in this regime was first studied by~\cite{Dubovsky:2015cca}), the frequency of which is given by 
\begin{equation}
	\nu  \simeq  \frac{n_e e^4 \log\Lambda}{2\pi m_e^2 v_e^3} \, ,
	\label{eq:Coulomb_collision}
\end{equation}
where $n_e$ and $m_e$ are the electron number density and mass, $v_e$ is the electron velocity in the rest frame of the proton, and $\Lambda$ is the Coulomb logarithm which takes on a value of $\log\Lambda \sim 10$ in plasmas of interest. The typical order of magnitude of $n_e$ near an astrophysical BH is discussed in Appendix~\ref{app:plasma}. To remain agnostic relative to the uncertainties of the ambient density and accretion flow, we shall assume $n_e\in(10^{-4},10^4)\,{\rm cm}^{-3}$. Correspondingly, the plasma frequency 
\begin{equation}
 \omega_p=\sqrt{\frac{4\pi n_e e^2}{m_e}}\approx 10^{-13}\left(\frac{n_e}{10^{-4}\,{\rm cm}^{-3}}\right)^{1/2} \,{\rm eV}\,, \label{omegap}
\end{equation}
ranges from $10^{-9}\,{\rm eV}$ to $10^{-13}\,{\rm eV}$.~\footnote{As originally recognized in Refs.~\cite{Pani:2013hpa,Conlon:2017hhi}, the plasma frequency can be in the mass scale for which superradiance is effective for astrophysical BHs in various mass ranges. We note, however, that in our case superradiance is not plasma-induced, but triggered by the bare mass of the dark photon.}

The characteristic timescale for electron-ion collisions
\begin{equation}
	\tau_{ei} = \nu^{-1} \sim \left( \frac{10^4 {\rm cm}^{-3}}{n_e} \right) \left(\frac{v_e}{10^{-3}}\right)^3 \, {\rm s} 
\end{equation}
should be compared with the oscillation period of the electric field,
\begin{equation}
 \frac{2\pi}{\omega_R} \sim 4\times 
10^{-4}\left(\frac{10^{-11}\,{\rm eV}}{m_{\gamma^\prime}}\right)\,{\rm s}\,,
\end{equation}
to determine whether collisions are important. In the absence of collisions (i.e., when $\tau_{ei}\gg 2\pi/\omega_R$), 
an oscillating electric field will induce oscillations in the plasma; the net work done on the plasma over a period of 
oscillation however will be zero, and thus energy in the field is not dissipated. As collisions become important (i.e., 
when $\tau_{ei}\lesssim 2\pi/\omega_R$), electrons and ions accelerated by the fields scatter prior to completing a 
full oscillation, dissipating energy in the process. Thus the expectation in the strong collisional regime is that the 
plasma may effectively absorb the dark photon field, prior to or during superradiance. 

Notice that for the largest number densities and smallest velocities considered in this work, collisions become 
important for dark photon masses $m_{\gamma^\prime} \lesssim 10^{-15}\, {\rm eV}$ and initial thermal velocities $v_{e} 
\simeq 10^{-3}\sqrt{T/(10^4\,\text{K})}$. However, if the electrons are accelerated by the (superradiantly grown) dark 
photon, the collisional rate will decrease dramatically and will become irrelevant for all dark photon masses of 
interest, given the strong dependence on $v_e$ of the Coulomb collision rate [Eq.~\eqref{eq:Coulomb_collision}]. As we 
will show in a moment, the electrons will be always accelerated up to relativistic speeds on the time scales of 
interest. Therefore we expect collisions could play an important role only in the initial stages of 
the superradiant growth. 

For small electron velocities (valid for the early stages of superradiance), one can drop both the magnetic field term as well as the non-linear gradient contribution to Eq.~\eqref{eq:vlasov}, yielding an electron velocity
\begin{equation}
	\vec{v}_e = \frac{-e}{m_e(\nu - i \omega)} (\vec{E} + \sin\chi_0 \, \vec{E^\prime}) \, .
\end{equation}
Writing the photon and hidden photon fields in Fourier modes, taking the non-relativistic limit, and decomposing the 
transverse and longitudinal field components yields two coupled sets of differential equations: 
\begin{eqnarray} \label{eq:mixingEq_1}
0 = \Big[(-\omega^2+k^2)  \hspace{.2cm} +& \\ 
	\begin{pmatrix}
		\frac{\omega_p^2} { (1 + i \nu / \omega)}&   \frac{\sin\chi_0 \omega_p^2 }{(1 + i \nu / \omega) }\nonumber\\
		 \frac{\sin\chi_0 \omega_p^2}{(1 + i \nu / \omega)} &\frac{ \sin^2\chi_0 \omega_p^2}{ (1 + i \nu / \omega)}  + m_{\gamma'}^2
	\end{pmatrix}\Big] \, &
	 \begin{pmatrix}
	 	A_T \\ A^\prime_T
	 \end{pmatrix}  
\end{eqnarray}
and 
\begin{eqnarray}\label{eq:mixingEq_2}
 0 = \Big[ -\omega^2 + 
	\begin{pmatrix}
	 \frac{\omega_p^2} { (1 + i \nu / \omega)}&  \frac{ \sin\chi_0 \omega_p^2} { (1 + i \nu / \omega)} \\
		 \frac{\sin\chi_0 \omega_p^2} { (1 + i \nu / \omega)} & \frac{ \sin^2\chi_0 \omega_p^2} { (1 + i \nu / \omega)} + \frac{m_{\gamma'}^2} {(1 - k^2 / \omega^2)}
	\end{pmatrix} \Big]
	 \begin{pmatrix}
	 	A_L \\ A^\prime_L
	 \end{pmatrix}  \, .\nonumber\\
\end{eqnarray}
The eigenvalues and eigenvectors of the mass matrix allow us to identify the mass and composition of the propagating states. 

The limit $m_{\gamma'} \gg \omega_p$ is trivial; in this case in-medium effects are not important and the eigenstates of 
the propagating mode with mass $m_{\gamma'}$ coincides with $A'$.

In the limit $m_{\gamma'} \ll \omega_p$ instead, the dark photon $A_{\rm dp}$ is identified as the propagating state with mass 
$m_{\gamma^\prime}$ and is comprised of a linear combination of $A$ and $A^\prime$, with 
\begin{eqnarray}
	A \sim - \sin\chi_0 \, A_{\rm dp}  \, \left(1 + \frac{{m_{\gamma'}}^2 (1 + i   \nu / {m_{\gamma'}}) }{\omega_p^2}\right) 
\\ 
	A^\prime \sim  A_{\rm dp}  \, \left(1 - \frac{{m_{\gamma'}}^2 \sin^2\chi_0 (1 + i  \nu / {m_{\gamma'}} )}{\omega_p^2} 
\right) \, .
\end{eqnarray}

The observable electric field is then given by the combination
\begin{equation}
	A_{\rm obs} = A + \sin\chi_0 \, A^\prime \sim -\sin\chi_0 \, \frac{i {m_{\gamma'}} \nu}{\omega_p^2} A_{\rm dp} \, .
\end{equation}
Note that $A_{\rm obs}=0$ when $\chi_0=0$, since we are focusing only on the dark electromagnetic field (which we shall 
occasionally refer to simply as the electromagnetic field, as the SM Maxwell field is irrelevant for our analysis).

We see here that an effective in-medium suppression of the mixing  proportional to 
$m_{\gamma^\prime} \nu / \omega_p^2$ arises at small masses. Physically, this happens because the plasma can 
efficiently move in response to 
the external field, oscillating in such a way so as to induce a partial cancellation. 

The frequency of the dark photon field can be obtained by solving the coupled differential equations expressed in 
Eqs.~\eqref{eq:mixingEq_1} and \eqref{eq:mixingEq_2}, and is in general complex. In the non-relativistic limit, the real 
part can be identified with the mass of the propagating state, while the imaginary part induces dissipation in the medium; this effect is akin to the concept of a skin-depth, where the imaginary part identifies the length scale over 
which the field drops by an $e$-fold~\cite{Dubovsky:2015cca}. The skin-depth of the dark photon field in the 
limit $m_{\gamma^\prime} \ll  \omega_p$ is given by
\begin{equation}
	\delta_{\rm dp, \, i} \sim  \left(\nu \frac{m_{\gamma^\prime}^2}{2 \omega_p^2} \sin^2\chi_0 \right)^{-1} \, .
\end{equation}
From this we see that the skin-depth of the dark photon is always much larger than any of the scales of interest. As a 
consequence, one can conclude that dissipation effects in the medium surrounding the BH are not sufficient to inhibit 
the growth of the dark photon field. Thus, the dark photon field will grow, and the (dark) electric field generated via 
this process will drive the electrons and ions to larger velocities until the electron-ion collision timescale has 
become longer than the characteristic driving frequency for all dark photon masses of interest. The immediate 
consequence is that the effect of collisions can be neglected. Therefore in the next section we will focus on the collisionless regime.

\subsection{Collisionless Regime}

Dropping the collision term in Eq.~\eqref{eq:vlasov} dramatically simplifies the problem at hand. In this case, one can 
derive a general solution for the response of the plasma to an oscillating transverse wave; we defer this derivation 
(generalizing previous electromagnetic solutions~\cite{doi:10.1063/1.1692942, 
Akhiezer1956THEORYOW,doi:10.1063/1.1692942,max1971strong} to include the presence of the dark photon) to 
Appendix~\ref{app:KawDawson}, and only present the results here. 

The mixing equation for the transverse modes in this case can be expressed as 
\begin{eqnarray} \label{eq:mixingEq_3}
0 = \Big[-\omega^2  + 
	\begin{pmatrix}
		\frac{\omega_p^2} {\gamma}&   \frac{\sin\chi_0 \omega_p^2 }{\gamma}\\
		 \frac{\sin\chi_0 \omega_p^2}{\gamma} &\frac{ \sin^2\chi_0 \omega_p^2}{\gamma}  + m^2
	\end{pmatrix} \Big]\,
	 \begin{pmatrix}
	 	A_T \\ A^\prime_T 
	 \end{pmatrix}   \, ,
\end{eqnarray}
where $\gamma$ is the time-averaged boost factor characterizing the motion of the plasma. In general, the plasma motion will be determined by the field itself, making this problem difficult to solve self-consistently (this is because the field is induced by the dark photon, which is a linear combination of $A$ and $A^\prime$, but the appropriate weights must be determined by solving the mixing equation, which itself depends on the weights). 

One can, however, identify various independent regimes in which these equations can be solved. During the initial stages 
of superradiance, the electric field induced by the dark photon cloud is small and the electrons are expected to be 
non-relativistic. In this regime, one can take either the limit in which $m_{\gamma^\prime} \gg \omega_p$ or $\omega_p 
\gg m_{\gamma^\prime}$, in each case finding 
\begin{equation}\label{eq:aobs_2}
	A_{\rm obs} \sim \begin{cases}
		\sin\chi_0 \, A_{\rm dp} \, \hspace{.9cm} {\rm if} \hspace {.3cm} m_{\gamma^\prime} \gg \omega_p \\
		\frac{m_{\gamma^\prime}^2}{\omega_p^2} \sin\chi_0 \, A_{\rm dp} \hspace{.3cm} {\rm if} \hspace {.3cm} m_{\gamma^\prime} \ll \omega_p \, .
	\end{cases}
\end{equation}
The former of these is consistent with recovering the vacuum mixing, while the latter experiences a strong in-medium 
suppression.  As the cloud continues to grow, the plasma may be driven to relativistic speeds. Should  $\omega_{p, {\rm 
eff}} \equiv \omega_p / \sqrt{\gamma} \ll m_{\gamma^\prime}$, one can see that Eq.~\eqref{eq:mixingEq_3} will reduce to 
the vacuum solution with $\sin\chi \sim \sin\chi_0$.  As shown in Appendix~\ref{app:KawDawson}, the $\gamma$ factor can 
be expressed in terms of the applied electric field as $\gamma = \sqrt{1+  (e E_{\rm obs} / (m_e \omega))^2}$.  Thus, 
for small dark photon masses $m_{\gamma^\prime}$ one must follow the evolution of the plasma boost factor as the 
superradiant cloud grows, assuming an observable electric field generated by the in-medium suppressed value show in 
Eq.~\eqref{eq:aobs_2}, to determine whether the boost becomes sufficiently large so as to remove the suppression 
altogether. In the numerical work that follows, we implement this transition using a sharp cut-off; while this is 
clearly naive, the transition is expected to occur rapidly and thus should represent a rough approximation to reality.

\subsection{Summary of In-Medium Suppression and Effect of Collisions}

To summarize, the various regimes discussed include
\begin{itemize}
	\item \emph{Vacuum regime:} $\nu, \omega_p \ll m_{\gamma^\prime}$. Here there are no in-medium effects. The external 
induced fields oscillate at a frequency larger than the natural oscillation frequency of the plasma, driving the plasma 
in such a way that no counterbalancing fields can be generated. Dissipative effects are strongly suppressed and can be 
neglected.
	\item \emph{Collisional regime:} $m_{\gamma^\prime} \ll \nu, \omega_p$. The dark photon interactions are initially 
suppressed by in-medium effects, and the large electron-ion collision rate induces an imaginary part of the dark photon 
frequency; this is equivalent to saying that the dark photon field dissipates energy continuously via inverse 
bremsstrahlung absorption as it propagates. For the environments of interest, this energy dissipation is never large 
enough to significantly suppress or absorb the dark photon field, and thus superradiance will occur and will drive $\nu 
\rightarrow 0$. Notice also that in general as $m_{\gamma^\prime} \rightarrow 0$ the dark photon fully decouples from 
electromagnetism. 
	\item \emph{Collision-less regime:} $\nu \ll m_{\gamma^\prime} \ll \omega_p$. For small field values, the dark photon will have an in-medium suppression of the mixing. However, as the observed electric field grows, the effective plasma mass will be driven toward zero and the vacuum mixing will be recovered. 
\end{itemize}
As we will show below, this in-medium suppression can be efficiently removed if the dark photon mass $m_{\gamma^\prime} \gtrsim 10^{-16}$ eV 
and the vacuum mixing $\sin\chi_0 \gtrsim 10^{-7}-10^{-8}$, depending on the ambient electron density.

\begin{table*}[t]
  \centering
  \renewcommand{\arraystretch}{1.5}
 \begin{tabular}{ |p{3cm}||p{7cm}|p{4cm}|}
 \hline
 Parameter & Description & Typical values of interest \\ [0.6ex] 
 \hline\hline
 $\chi$ & Kinetic mixing & $10^{-8}-10^{-2}$ \\ 
 \hline
 $m_{\gamma'}$ & Dark photon mass & $10^{-18}-10^{-10} \rm eV$ \\
 \hline
 $n_e$ & Electron density & $10^{-4}- 10^{4} \rm cm^{-3}$\\
 \hline
 $\omega_p=\sqrt{(4\pi \alpha n_e) / m_e}$ & Plasma frequency & $10^{-13}-10^{-9} \rm eV$ \\
 \hline
 $n_{\gamma'}^{\rm sr}$ & Saturation number density & lim$_{\chi \rightarrow 0}$  $10^{57} \rm cm^{-3} \frac{m_{\gamma'}}{10^{-11}\rm eV}$\\
 \hline
 $\left< \gamma \right>$ & Boost factor & $1 - 10^{13}$ \\
 \hline
 $n_{\gamma'}^{\rm ionization}$ & Number density requited to ionize hydrogen & $10^{34} \frac{1}{\sin^2\chi} \frac{10^{-11}\rm eV}{m_{\gamma'}} \rm cm^{-3}$ \\
 \hline
  \end{tabular}
   \caption{ Key parameters determining if and when dark photon superrradiance quenches. These parameters are: the kinetic mixing $\chi$ and the mass $m_{\gamma'}$ of the dark photon,  $n_e$ the electron number density near the BH, the plasma frequency $\omega_p$, the dark photon number density when superradiance begins extracting a significant fraction of the BH spin (equivalently this can be thought of the the approximate upper limit on the dark photon number density in the limit that the boson is truly non-interacting), time-averaged boost factor $\left< \gamma \right>$, and the dark photon number density when neutron hydrogen ionizes $n_{\gamma'}^{\rm ionization}$.}
  \label{tab:1}
\end{table*}

\section{Quenching the Growth of the Dark Photon} \label{sec:dps}

In this section we outline the dominant quenching mechanisms for kinetically mixed dark photons. We show that for 
sufficiently large kinetic mixings, dark photon superradiance may produce time-oscillating electromagnetic signatures 
arising from semi-Compton scattering and synchrotron emission of the ambient electrons. In Table~\ref{tab:1} we provide a glossary for a number of fundamental parameters which will control the superradiance quenching (see also Fig.~\ref{fig:evolution} for an illustration of the growth and quenching process). 

\subsection{Scattering Processes}

The evolution and quenching of the superradiant instability in the context of the SM photon was recently studied 
in-depth in~\cite{Cardoso:2020nst,Blas:2020kaa}; the case of the dark photon can be understood analogously with two 
notable 
exceptions. First, unlike the SM photon, the dark photon has a bare mass that will not be modified by the presence of 
strong electric fields or by modifications to the local electron density. Second, the dark photon scattering rate 
intrinsically depends on the properties of the ambient plasma, which themselves may depend on the energy density of the 
dark photon in a more involved way than for the SM photon. 
In both cases, however, one might expect  the exponentially growing boson cloud to generate strong (dark) electromagnetic fields, and if these relativistic oscillations can be reached, Compton scattering (or more 
appropriately semi-Compton in the case of the dark photon) and synchrotron emission may produce energy losses capable of 
balancing the energy being extracted from the BH spin. The focus on what follows is on identifying the dark photon 
parameter space for which this occurs. It is important to note that we do not require the dark photon to be the dark 
matter, as the superradiant instability can be triggered for arbitrarily small abundances.

Owning to the kinetic mixing, the dark photon will induce an electric force on the ambient electrons that oscillates with frequency $\omega_R$. Assuming the motion of the electrons to be dominated by that of the electric field (a valid assumption at large number densities), one can approximate  the time-dependent boost factor of the electrons as
\begin{equation}
	\gamma(t) = \sqrt{1 + \sin^2\chi \, \frac{n_{\gamma^\prime}\, \cos^2(\omega_R t)}{m_e^2 m_{\gamma^\prime}} } \, .
\end{equation}
In the limit of large dark photon number density, the time-averaged gamma factor is approximately given by $\left< \gamma \right> \sim \sin\chi \, \sqrt{n_{\gamma^\prime}/ m_{\gamma^\prime}} / m_e \simeq 5.4 \times 10^3 \sin\chi \left(\frac{n_{\gamma'}}{10^{21}\rm cm^{-3}}\right)^{1/2}\left(\frac{10^{-12} \rm eV}{m_{\gamma'}}\right)^{1/2}$. In this case the energy loss from semi-Compton scattering $\gamma' + e^- \leftrightarrow \gamma + e^-$ is given by

\begin{align}
\frac{dE_{\rm sC}}{dt} &= \frac{4}{3}(\gamma^2 - 1) \sin^2\chi \, \sigma_T \rho_{\gamma^\prime} \, n_e \nonumber \\ 
 & \stackrel{\gamma \gg 1}{\sim} \frac{4}{3}\sin^4\chi \frac{n_{\gamma^\prime}^2}{m_e^2} \sigma_T n_e \, .
\end{align}
Similarly, the energy loss rate via synchrotron emission, is given by 
\begin{align}
	\frac{dE_{\rm syn}}{dt} &= \frac{dE_{\rm sC}}{dt} \sin^2\zeta \, ,
\end{align} 
where $\zeta$ is the angle between the electron velocity and the induced magnetic field. We have verified explicitly 
that the typical center of mass energy of the photons $\sim \gamma m_{\gamma^\prime}$ is never large enough that 
Klein-Nishina corrections become important.

\subsection{The Evolution and Quenching}

In order for the above processes to be capable of quenching the growth of the dark photon cloud, they must be faster than superradiant itself, which injects energy at a rate 
\begin{equation}
	\frac{dE_{\rm sr}}{dt} \simeq 2 \frac{m_{\gamma^\prime} \, n_{\gamma^\prime}}{\tau_{\rm sr}} \, .
\end{equation}

The dark-photon condensate will grow until the two energy losses compensate each other, which happens at the saturation 
particle density (for $m=1$, and assuming the plasma is already relativistic)
\begin{align}
 n_{\gamma^\prime}^{\rm sat} &\sim \frac{6 m_e^2 \tilde{a}}{n_e\sigma_T \sin^4\chi} m_{\gamma^\prime}^2 \alpha^6 \nonumber\\
 &\approx 10^{21}\,{\rm cm}^{-3}\frac{\tilde{a}}{\sin^4\chi}\left(\frac{10^4\,{\rm cm}^{-3}}{n_e}\right)\left(\frac{m_{\gamma^\prime}}{10^{-12}\,{\rm eV}}\right)^2\left(\frac{\alpha}{0.1}\right)^6 \,, \label{nsaturation}
\end{align}
where $\tilde a$ is the initial dimensionless spin of the BH.
At saturation the field has extracted from the BH an amount of energy given by
\begin{equation}
 M_{\rm cloud} = m_{\gamma^\prime} n_{\gamma^\prime}^{\rm sat} {\cal V}\,,
\end{equation}
where ${\cal V}=\frac{4\pi}{3}r_{\rm cloud}^3$ is the volume of the cloud. For a single azimuthal mode (typically $m=1$, see Ref.~\cite{Ficarra:2018rfu} for the case of multiple modes), any energy extraction $\delta M$ is proportional to the angular momentum extraction $\delta J$ through $\delta M=\omega_R/m \delta J$~\cite{Brito:2015oca}. 

\begin{figure}
\centering
\includegraphics[width=0.49\textwidth]{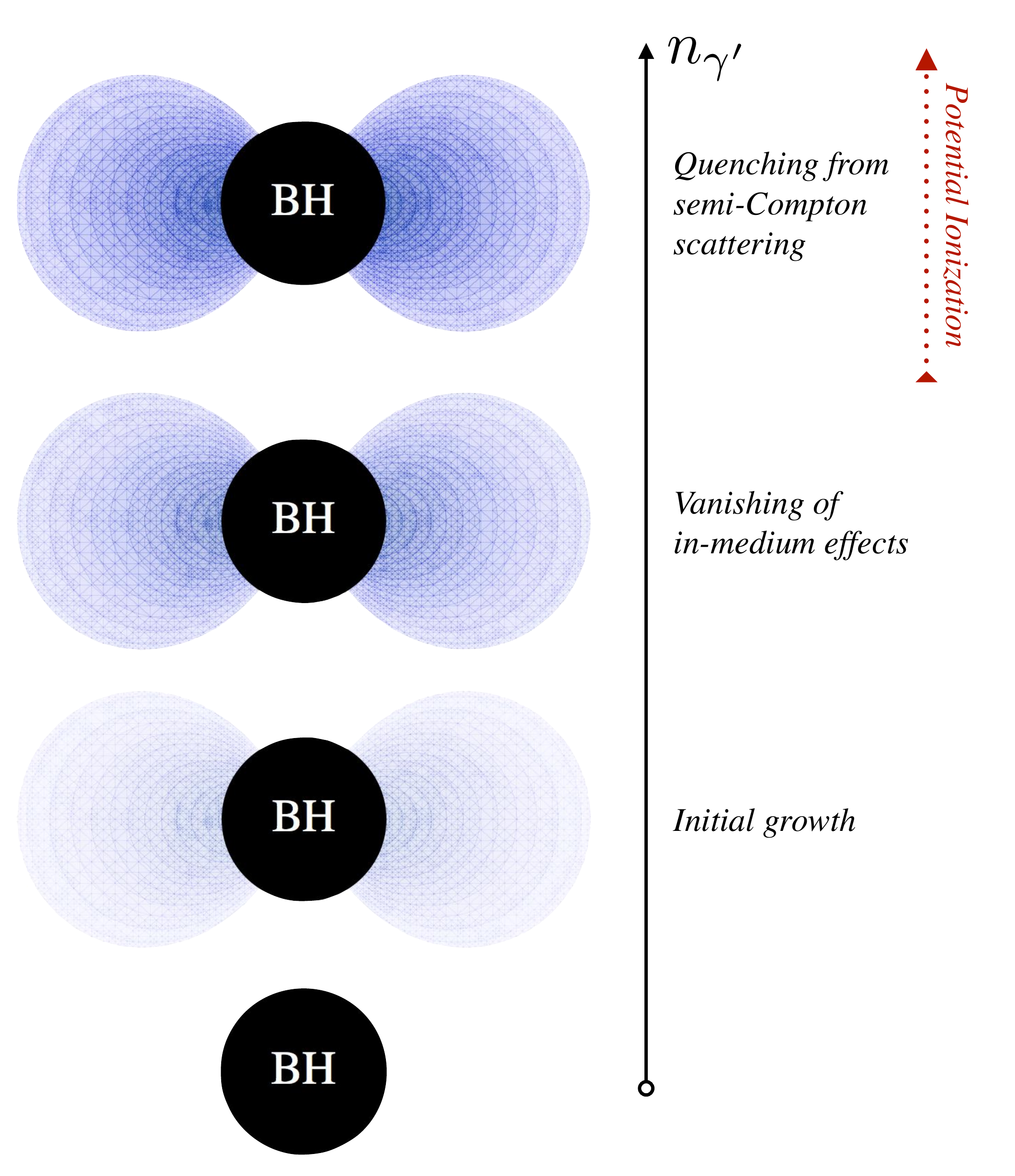}
\caption{\label{fig:evolution}  Heuristic evolution illustrating the growth and quenching of the superradiant dark photon cloud. For dark photons with vacuum kinetic mixings $\chi_0 \gtrsim \chi_{\rm min}$ (\ie the minimum mixing required to quench superradiance) this behavior can be characterized as follows: after an initial stage of growth, the in-medium suppression of the mixing (should it exist) is removed due to the large electric fields generated by the dark photon. At larger number densities the growth fully saturates as semi-Compton scattering removes energy at a rate equivalent to the energy injection rate from superradiance. Depending on the dark photon mass, the Stark effect may ionize all neutral hydrogen atoms in the latter stages near saturation. }
\end{figure}

In order to evade current bounds coming from gaps in the BH mass-spin plane or from gravitational-wave emission, we 
require that quenching is sufficiently efficient so that it would allow for extraction of a negligible amount of 
angular momentum, i.e. $\delta J/J\ll1$. Using the equations above, this is equivalent to
\begin{equation}
 \sin^4\chi \gg \frac{8\pi m_e^2 }{n_e\sigma_T} m_{\gamma^\prime}\alpha\,.
\end{equation}
Note that, as previously discussed, $\chi$ is the in-medium coupling which depends on various parameters of the system, including the number density of dark photons (which controls whether the in-medium suppression has been removed). 
In the next section we shall use this condition to identify the parameter space in the $\sin\chi_0-m_{\gamma^\prime}$ plane where the superradiant instability is efficiently quenched; we heuristically outline the growth and saturation of the cloud for a generic dark photon model in Fig.~\ref{fig:evolution}.

In order to determine whether a given dark photon model, defined by mass and vacuum mixing angle, quenches before 
extracting sufficient BH spin, one must first define both the BH mass of interest, or equivalently the superradiant coupling
$\alpha \equiv m_{\gamma^\prime} M$, and the local electron number density. The former is responsible for setting the 
superradiant time scale, and the latter controls the efficiency of the semi-Compton and synchrotron emission. In order 
to be conservative, we adopt a fiducial value of $\alpha = 0.4$, for which $\tau_{\rm sr}$  of a near-extremal BH is 
maximal, however comparable values of $\alpha$ do not yield significantly different results. The range of $n_e$ expected in the vicinity of a BH is estimated in Appendix~\ref{app:plasma} and
may span many orders of magnitude. Rather than adopting a particular value, in what follows we show 
results under various assumptions in order to be as broad as possible.

A final word is in order regarding the electron number density. The ambient gas around BHs after reionization or in environments with large thermal accretion rates is expected to be largely ionized. This may not necessarily be the case however, implying the free electron number density may be suppressed relative to that of neutral hydrogen. This opens the possibility that the neutral hydrogen can be ionized during the superradiant growth, further increasingly the electron number density, and subsequently the scattering rate. Ionization from the growing dark photons can occur via the Stark effect~\cite{Blas:2020kaa}, in which the induced electric field becomes large enough to shift the ground state of neutral hydrogen from $13.6$ eV $\rightarrow 0 \, $eV. Following \cite{Blas:2020kaa}, the number density of dark photons required to ionize the hydrogen is 
\begin{equation}
	n_{\gamma^\prime}> n_{\gamma^\prime}^{\rm ionization} \sim 10^{35}\, \frac{1}{\sin^2\chi}\left(\frac{10^{-12} \, {\rm eV}}{m_{\gamma^\prime}} \right) \, {\rm cm}^{-3} \, ,
\end{equation}
which should be compared to Eq.~\eqref{nsaturation}, suggesting this this may happen when
\begin{equation}
 \sin^2\chi\lesssim 2\times 10^{-14} \tilde{a}\left(\frac{10^4\,{\rm cm}^{-3}}{n_e}\right)\left(\frac{m_{\gamma^\prime}}{10^{-12}\,{\rm eV}}\right)^3\left(\frac{\alpha}{0.1}\right)^6\,. \label{stark}
\end{equation}
The above equation assumes $n_{\gamma^\prime}^{\rm ionization}$ is still sufficiently low so that the black hole has not yet been spun down.

\begin{figure}
	\centering
	\includegraphics[width=0.49\textwidth]{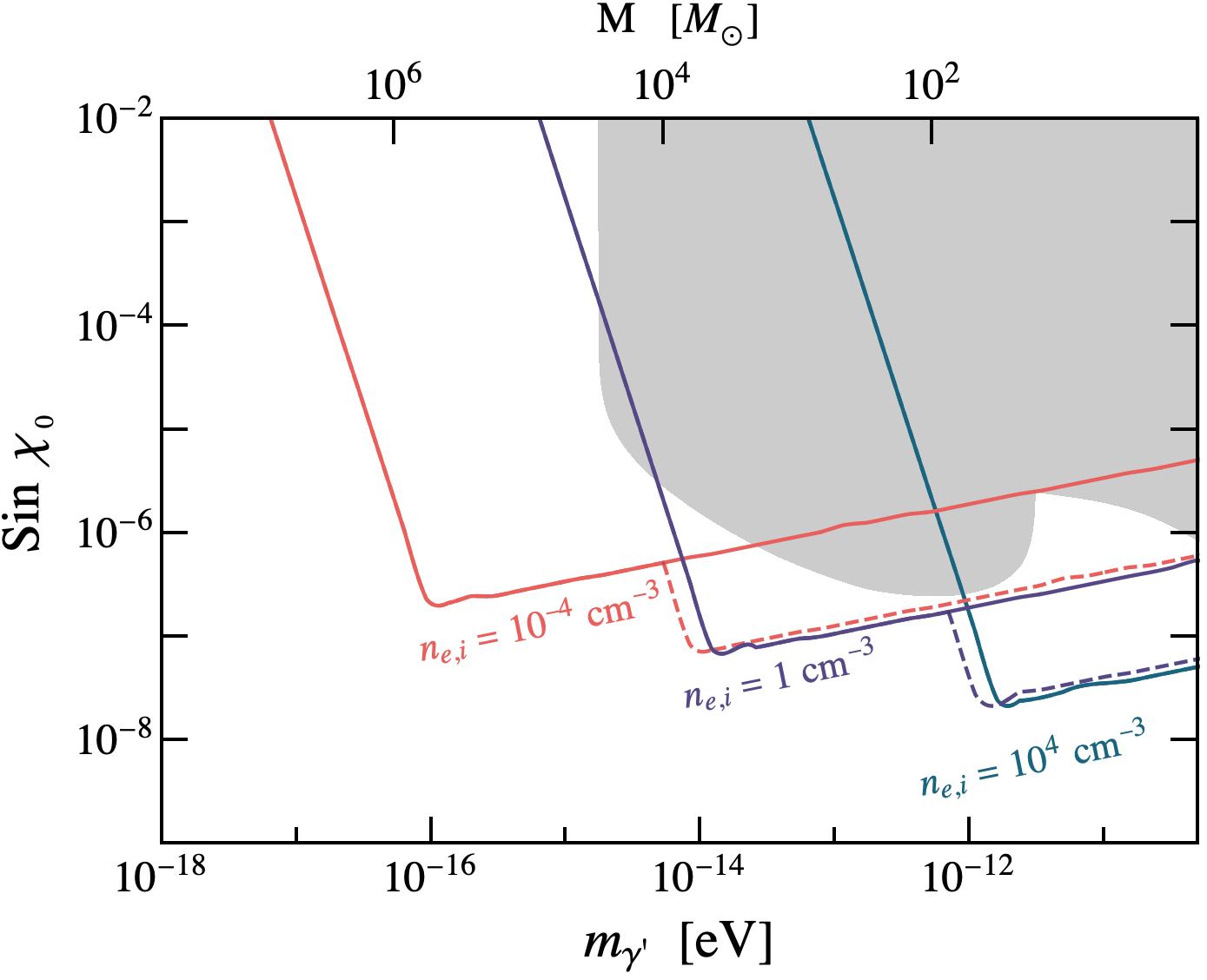}
	\caption{\label{fig:kin_mix}  Lower bound on kinetic mixing $\chi_0$ for which quenching of superradiance occurs, assuming local electron number densities of $10^{-4}$, $1$, and $10^4 \, {\rm cm}^{-3}$ and $\alpha = 0.4$. Short dashed lines account for the possibility that the Stark effect could increase the local electron number density if the initial medium is not fully ionized (see text for details; $x_e$ taken to be $2 \times 10^{-4}$). The approximate range of BH masses being probed is shown on the upper x-axis, and is derived assuming $\alpha = 0.4$. Kinetic mixings below the values shown here may still be capable of spinning down massive BHs, depending on additional details of the model. The gray region corresponds to current constraints on dark photons, derived from the cosmic microwave background~\cite{Caputo:2020rnx,Caputo:2020bdy}.}
\end{figure}

The minimum vacuum kinetic mixing $\chi_0$ capable of quenching superradiance at each dark photon mass is shown in Fig.~\ref{fig:kin_mix} along with current constraints on dark photons, derived from the cosmic microwave background~\cite{Caputo:2020rnx,Caputo:2020bdy} (note that these constraints are derived assuming dark photons do not constitute dark matter; in the case that they do comprise a non-negligible fraction, the constraints are many orders of magnitude stronger~\cite{McDermott:2019lch,Witte:2020rvb, Caputo:2020bdy}). Here, we show results for $\alpha = 0.4$ and $n_e = 10^{-4} \,{\rm cm}^{-3}$  (red), $1 \, {\rm cm^{-3}}$ (purple), and $10^4 \,{\rm cm}^{-3}$  (blue). We also show the impact of adopting a local free electron fraction $x_e \equiv n_e / n_H = 2 \times 10^{-4}$ (consistent with the mean value of the Universe prior to reionization) -- at larger masses, the local hydrogen can ionize via the Stark effect and superradiance can be quenched for smaller kinetic mixings. 
The reduced sensitivity at low masses arises because the in-medium suppression of the kinetic mixing cannot be removed.

The overall physical trend is manifest: an increase in the electron number density can be compensated for by adopting a 
smaller mixing angle.   We also note that a smaller superradiance parameter $\alpha$ would imply a smaller growth rate for the 
superradiant cloud $\frac{dE_{\rm sr}}{dt}$,  thus allowing smaller scattering rates (achieved \eg by lowering the mixing 
angle) to induce quenching. Low dark photon masses are notoriously difficult to probe due to the strong in-medium suppression; nevertheless, it seems quenching can be achieved to some degree for masses as small as $\sim 10^{-17}$ eV.

\subsection{Observable signatures}

We have shown in the previous section that dark photons with vacuum mixings $\chi_0 \gtrsim 10^{-8}$, depending on the mass, can quench superradiance prior to the spin down of the BH. This quenching is a consequence of the fact that the accelerated electrons radiate synchrotron emission and semi-Compton scatter off the ambient dark photon. These processes result in the direct emission of photons, with the characteristic energy $\left< E_c \right>$ of each processes roughly given by the inverse semi-Compton formulas $\sim\gamma^2 \omega_g$ and $\sim\gamma^2 m_{\gamma^\prime}$, respectively, where $\omega_g = e B / (m_e \gamma) $ is the gyro frequency.

\begin{figure*}
	\centering
	\includegraphics[width=0.45\textwidth]{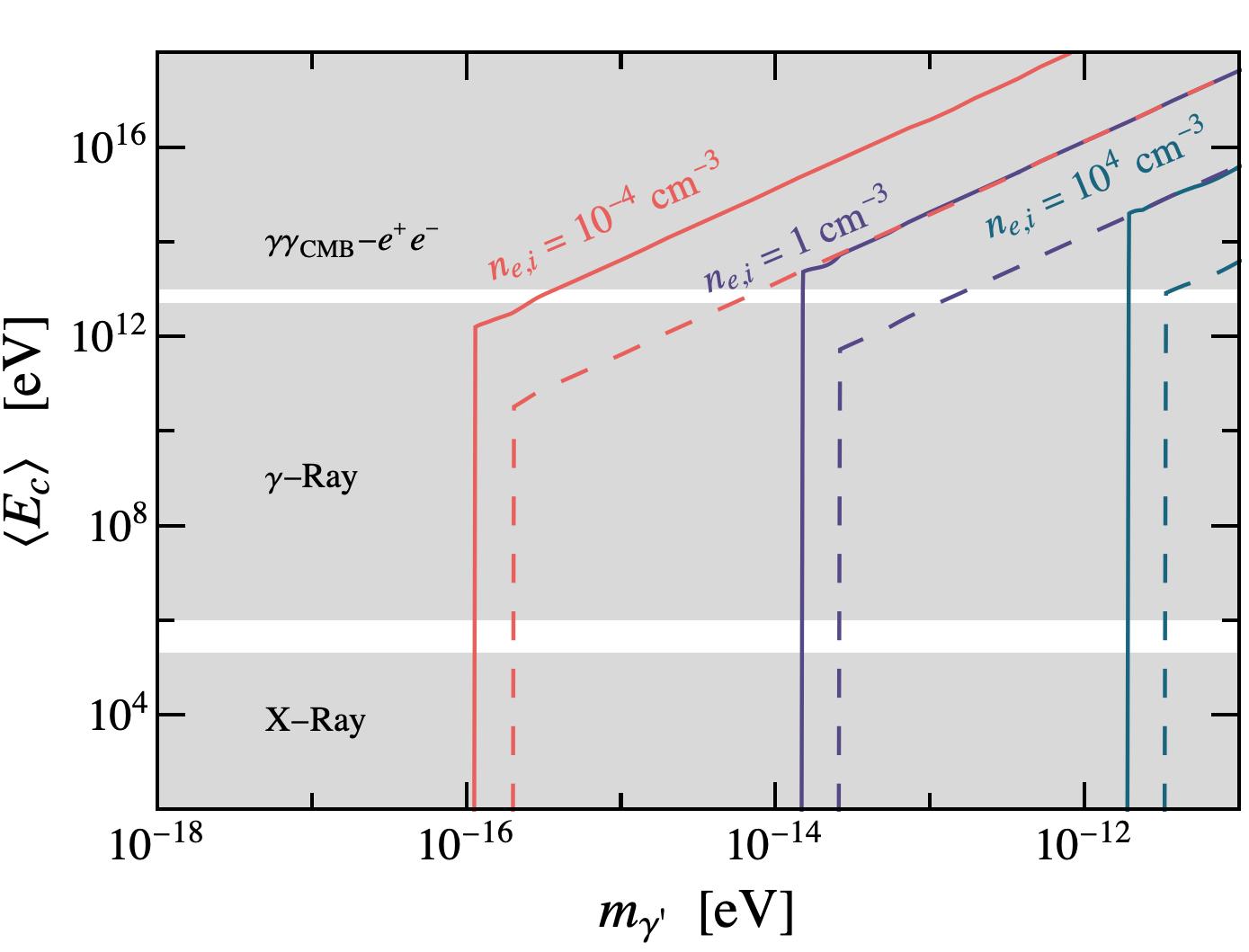} \hspace{.3cm}
 \includegraphics[width=0.48\textwidth]{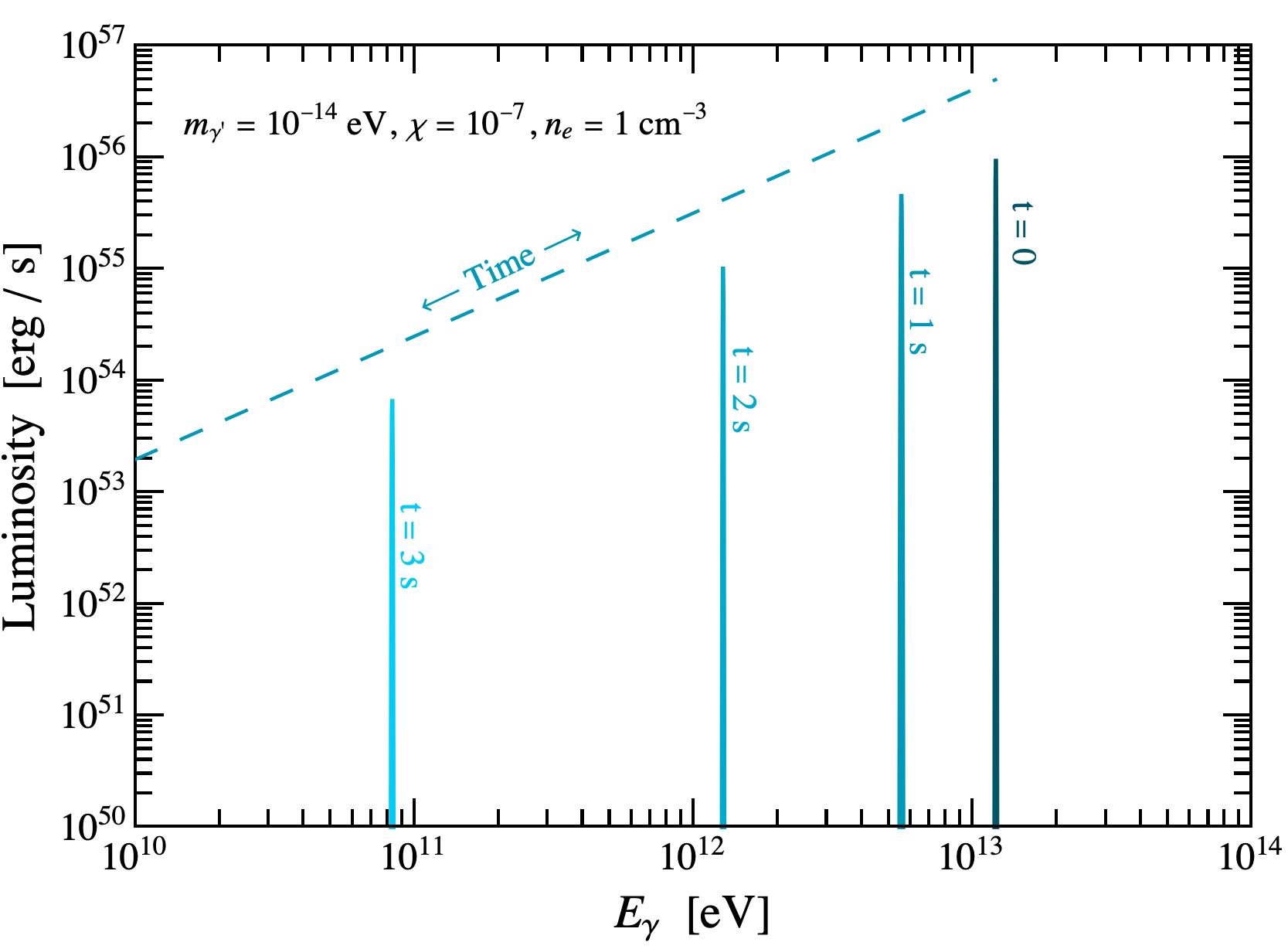}
	\caption{\label{fig:lumi} {\bf Left:} Characteristic time averaged energy $\left<E_c \right>$ of up-scattered photons, given by $\gamma^2 m_\gamma$, as a function of dark photon mass, assuming various values of the local electron number density and for $M m_{\gamma^\prime} = 0.4$. Solid lines take $\chi_0$ to be the minimum value capable of quenching superradiance $\chi_{\rm min}$, while dashed blue line takes $\chi_0 = 10 \chi_{\rm min}$ for one representative case. Approximate energy ranges over which telescopes observe $\gamma$-rays, $x-$rays, and the CMB are highlighted. We also highlight the regime where $\gamma-$rays can scatter of CMB photons and pair produce $e^+e^-$; this process may generate TeV halos in a similar mechanism to blazars. {\bf Right:} Time evolution of the electromagnetic signal for $m_{\gamma^\prime} = 10^{-14} \, {\rm eV}$, $\sin\chi_0 = 10^{-7}$, and $M m_{\gamma^\prime} = 0.4$. }
\end{figure*}

The luminosity of each process is roughly given by $\mathcal{V} \times dE/dt$ at the saturation density~\eqref{nsaturation}. Assuming $\chi$ is  sufficiently large to saturate the growth, the maximum luminosity and characteristic energy for the semi-Compton process 
then yields  
\begin{eqnarray}
    L^{\rm max}_{\rm sC} \simeq 3\times 10^{58}\frac{\rm erg}{\rm s} \tilde{a}^2\Big(\frac{m_{\gamma'}}{10^{-12}\text{eV}}\Big)\Big(\frac{\alpha}{0.4}\Big)^9 \nonumber \\ \times \Big(\frac{10^{-8}}{\sin\chi}\Big)^4\Big(\frac{10^4 \text{cm}^{-3}}{n_e}\Big), \label{eq:LumiMax}
      \\
    E^{\rm max}_{\rm sC} \simeq 2.1\times 10^{15} \text{eV} \, \tilde{a}\Big(\frac{\alpha}{0.4}\Big)^6 \times \Big(\frac{10^{-8}}{\sin\chi}\Big)^2  \nonumber \\ \times \Big(\frac{10^4 \text{cm}^{-3}}{n_e}\Big) \Big(\frac{m_{\gamma'}}{10^{-12}\text{eV}}\Big)^2\, ,
    \label{eq:EMax}
\end{eqnarray}
while in the case of synchrotron emission one finds
\begin{eqnarray}
	 L^{\rm max}_{\rm syn} \simeq \sin^2\xi L_{\rm sC}^{\rm max} \,,\\
	 E^{\rm max}_{\rm syn} \simeq  \gamma \frac{e B}{m_e} \simeq \frac{\sin\chi e}{m_e^2} n_{\gamma'} \, .
	 \label{eq:MaxSyn}
	 \end{eqnarray}
As expected, these are comparable processes, with semi-Compton dominating due to the lack of dependence on orientation. We also stress that Eqs.~\eqref{eq:LumiMax}-\eqref{eq:MaxSyn} are valid \emph{only if} superradiance is quenched. In particular, the maximum luminosity reached can never exceed the upper limit set by the physical scales in the problem $\sim M/\tau_{s}$, which is always well below Planckian luminosities~\cite{Cardoso:2018nkg}.

We illustrate the dependence of the  characteristic energy of the semi-Compton up-scattered photon $\left<E_c \right>$ 
(left) and  time-averaged luminosity arising from the time-dependent in the gamma factor (right) in 
Fig.~\ref{fig:lumi}. Results are shown by either fixing the kinetic mixing to the minimal value $\chi_{\rm min}$ 
required for quenching (solid), or by taking $\chi_0 = 10 \times \chi_{\rm min}$ (dashed), and for various electron 
number densities. The luminosity is maximal for the smallest kinetic mixings (because the latter correspond to higher 
saturation density of the dark-photon cloud), reaching values near $\sim 10^{57} $ erg /s, well above the luminosities 
of \eg typical AGN. We have also highlighted in  Fig.~\ref{fig:lumi} various parts of the electromagnetic spectrum where 
this signal may be easily detectable, as well as the approximate threshold where $e^+e^-$ pair production may become 
important (see text below).  Intriguingly, because the electron boost factors vary with time (due to the oscillating 
electric fields), both $E_c$ and the luminosity have a characteristic oscillatory behavior; this is roughly illustrated 
in the right panel of Fig.~\ref{fig:lumi}, where we show the evolution over part of a period of oscillation for a 
particular dark photon candidate. Notice in this figure we have drawn the fixed time emission as a narrow line 
(importantly, this is an illustration, not a calculation); this is because both the dark photon and the plasma have 
narrow distributions in energy, the former being non-relativistic and the latter being driven by the electric field. The 
line width is likely generated by non-linearities in the dark photon field; while it is of great interest to understand 
the spectral characteristics, we consider determining this feature to be beyond the scope of this work.

Given that these (ordinary) photons are produced near the BH itself, a logical question to ask is whether they may 
escape their local environment. Assuming the local gas is ionized, the dominant interaction for photons with energies 
$E \lesssim 10^{7}$ eV is again that of Compton scattering. In this case, the optical depth can be calculated via 
\begin{equation}
	\tau = \int d\ell n_e \sigma_T  \, .
\end{equation}
If we parameterize the electron number density with a Bondi accretion profile, \ie $n_e \propto r^{-3/2}$ (see Appendix~\ref{app:plasma}), this integral is dominated by the highest densities near the BH itself. Approximating the lower limit of integration as the Schwarzschild $r_s$, we can write the optical depth as
\begin{equation}
	\tau \sim 10^{-7} \, \left(\frac{M}{10^8 M_\odot} \right) \, \left( \frac{n_{e, c}}{10^4 \, {\rm cm ^{-3}}}\right) \, ,
\end{equation}
where $n_{e,c}$ is the central electron number density at $r_s$.

For higher energy gamma rays, the optical depth will be set either by the possibility for photons to pair produce 
$e^+e^-$ by scattering off the extragalactic background light (EBL) or via $\gamma + p \rightarrow p + e^+ + e^-$. The 
cross section for the former is given by~\cite{Franceschini:2008tp,vestrand1983gamma,2011hea..book.....L} 
\begin{equation}
	\sigma_{\gamma\gamma} = \frac{3\sigma_T}{16}(1-\beta^2)(2\beta(\beta^2-2)+ (3- \beta^4) \log\left(\frac{1+\beta}{1-\beta}\right)) \, ,
\end{equation}
where $\beta = \sqrt{1 - 4 m_e^2 / s}$, and $s = 2 E_{\gamma} E_{\gamma_{EBL}} (1 - \cos\theta)$. The mean free path for this can be expressed roughly as~\cite{neronov2010evidence}
\begin{equation}
	\lambda_{\gamma\gamma} \sim 80 \left( \frac{E_\gamma}{10 \, {\rm TeV}} \right)^{-1} \, {\rm Mpc} \, .
\end{equation}
The scattering off ambient protons has a cross section~\cite{vestrand1983gamma}
\begin{equation}
	\sigma_{\gamma p} = \alpha r_0^2 \left[\frac{28}{9} \log\left( \frac{2 E_\gamma}{m_e}\right)  - \frac{218}{27}\right] \, ,
\end{equation}
where $r_0$ is the classical electron radius. Again assuming a fully ionized medium and the radial taken from Bondi 
accretion, one can compute the optical depth; we find this the optical depth for this process is always quite small and 
can be neglected. Thus, depending on the distance to the BH of interest, the $\gamma\gamma$ scattering rate may be 
sufficient to absorb the high energy photons. In this case, the electrons produced could scatter off ambient light and 
generate $\gamma$-ray halos around the BH in a similar manner to that of blazars~\cite{neronov2010evidence}. The 
detailed observational signatures are certainly worth investigating, and something we leave to future work.

\subsection{Comments on UV completions}

In the previous section we adopted a phenomenological approach and simply assumed that a dark photon with a given mass 
and kinetic mixing exists. It is however interesting to understand whether the quenching mechanism can depend on the 
origin of the dark photon mass. One can generically consider two different scenarios: either the mass is generated via 
the St\"uckelberg mechanism, which can be obtained from the Proca Lagrangian by introducing a fictitious gauge 
symmetry~\cite{Stueckelberg:1900zz}, or from the Higgs mechanism with a new scalar field $\Phi$ acquiring a vacuum 
expectation value $\left<\Phi \right> = v / \sqrt{2}$. 

The case in which the dark photon mass is generated via the St\"uckelberg mechanism leads to no novel quenching 
mechanisms, as no new interactions have been introduced. It is worth mentioning, however, that the so-called swampland 
conjecture, which is an attempt at identifying the set of effective field theories consistent with a theory of quantum 
gravity, presents a theoretical challenge for light dark photons with masses generated in this way~\cite{Reece:2018zvv}. 
Specifically, it has been argued that, for a given mass and mixing, one can compute the cut-off of the effective 
theory -- in order for this theory to remain meaningful, the cut-off scale should likely be $\gtrsim 
\mathcal{O}(1)\,{\rm TeV}$, implying kinetic mixings much stronger than those considered here.

In the case of the Higgs mechanism the situation may be substantially different~\cite{Fukuda:2019ewf}. Consider the following Lagrangian
\begin{equation}
    \mathcal{L} = -\frac{1}{4}F'_{\mu\nu}F'^{\mu\nu} + \frac{1}{2} |D_{\mu}\Phi|^2-V(\Phi),
\end{equation}
with $D_{\mu} = \partial_{\mu} - i g A'_{\mu}$ being the covariant derivative and $\Phi$ being the generic scalar field which is then assumed to acquire a vev, so that in the unitary gauge  the field may be expanded as $\Phi = (v + \sigma)/\sqrt{2}$. In order to be concrete, we consider here a  potential of the form $V(\Phi) = \lambda (|\Phi|^2 - v^2/2)^2$.

After symmetry breaking, the dark photon acquires a mass $m_{A'} = g v$; the mass of the scalar field can also be read 
from the potential as $m_{\sigma} = \sqrt{2\lambda} v$. Notice that when the vector field is amplified around the 
rotating BH, the scalar field picks an extra term due to the large densities $V(\Phi) \supset 
\frac{1}{2}g^2(v+\sigma)^2 \times \left< A_{\mu}A^{\mu}\right>$. One can check that in this case if 
$\left<A_{\mu}A^{\mu}\right>  > \frac{\lambda v^2}{g^2}$, the original symmetry is restored, and subsequently the dark 
photons become massless and can free stream away from the BH. However, unless one chooses extremely small  \xout{and 
unnatural} values of the scalar vev~\footnote{Note that for very small dark photon and dark Higgs  masses the semi-Compton 
scattering will not be suppressed by in-medium effect, as in this regime the dark photon would couple to SM fermion as 
in a millicharge model~\cite{Ahlers:2008qc, An:2013yua}. However, the interaction will anyway be likely suppressed by the 
smallness of the required dark gauge coupling.} and self-interaction $\lambda$, the symmetry restoration scale is never 
achieved for the dark photon masses considered. It is also worth mentioning that the operator $|\Phi|^2$ would also 
couple to any neutral operator built out of SM fields -- should the new vev be too low, one may also have to fine tune 
these additional couplings around zero to be consistent with current observations.

Finally, it is worth mentioning that the existence of any light particle in a theory directly coupled to the dark 
sector will allow for Schwinger pair production at sufficiently high densities~\cite{Schwinger:1951nm}. This may be 
particularly relevant for models in which the dark photon serves as a mediator for other dark fermions playing the role, 
for example, of dark matter~\cite{Pospelov:2008zw, Pospelov:2007mp, Dutra:2018gmv}.

In the case of the SM photon, the Schwinger pair production rate is given by~\cite{PhysRev.82.664}
\begin{equation}
	\Gamma_{\rm Schw} \simeq \frac{m_e^4}{4\pi^3} \left(\frac{\mathcal{E}}{\mathcal{E}_c} \right) \, \sum_{b=1}^\infty \frac{1}{b^2} \, e^{- b\pi \frac{\mathcal{E}_c}{\mathcal{E}}} \, ,
\end{equation}
where $\mathcal{E}_c = m_e^2 / \sqrt{4\pi\alpha}$ and $\mathcal{E}$ is the electric field strength. 
Analogously, given a dark sector fermion with mass $m_d$ -- charged under the $U(1)_{\rm dark}$ -- with Lagrangian 
$\mathcal{L}_{\rm dark} \supset \lambda_d \bar{\psi}_d \gamma^{\mu} \psi_d A^\prime_{\mu}$, the exponential suppression in 
the particle production rate will be removed when the superradiant cloud has reached a number density 
\begin{equation}
	n_{\gamma^\prime}^{\rm Schw} \sim \frac{\pi^2 \, m_d^4}{\lambda_d m_{\gamma^\prime}} \, .
\end{equation}
Comparing this value to the maximum achievable number density by superradiance, one finds that in order to quench superradiance the mass of the dark sector particle must satisfy
\begin{equation}
	m_d \lesssim 6.7\times10^6 \left(\frac{\lambda_d}{0.1}\right)^{1/4} \left( \frac{m_{\gamma^\prime}}{10^{-12} \, {\rm eV}}\right)^{1/2} \left( \frac{\alpha}{0.4}\right) ^{5/4} \, {\rm eV} \, .
\end{equation}

\section{A Comment on the SM Photon mass}\label{sec:sm}

Despite the fact that the SM photon is generically considered to be massless, there still exists the possibility that it carries a non-zero bare mass; if this is the case, it could shed light on many fundamental puzzles such as why charge is quantized and the potential existence of charged BHs~\cite{Adelberger:2003qx}. Additionally, in the contest of bosonic strings, the condensation of tensor fields generates a violation of Lorentz symmetry and may dress the photon with a small mass~\cite{PhysRevD.39.683, Bonetti:2016vrq}. In view of these intriguing theoretical possibilities, it is important to robustly constrain the magnitude of the photon mass. Typical probes to date involve: laboratory experiments using high-frequency tests of Coulomb's law~\cite{PhysRevLett.26.721}, study of the sector structure of the Solar wind in the presence of a finite photon mass~\cite{Ryutov:1997zz,Retin__2016}, and frequency-dependent dispersion in Fast Radio Bursts (FRB)~\cite{Bonetti:2017pym}.  In fact, if the photon has a bare mass $m_{\gamma}$, it may trigger a superradiant instability in the same way as the dark photon, and the observation of highly spinning supermassive BHs has been used as an argument to constrain bare photon masses above  $\sim 10^{-20} \,$ eV  ~\cite{Pani:2012vp}.

It is important however to stress that the propagating mode, exciting the superradiance instability, will have an energy 
gap in the dispersion relation coming from both the Proca mass and the plasma contribution. Therefore, for low photon 
masses $m_{\gamma} \lesssim 10^{-13}$ eV (note that current observations of FRBs constrain bare mass contributions above this level~\cite{Bonetti:2017pym}), the plasma contribution will always be the dominant one, even for low density 
environments, see Eq.~\eqref{omegap}. This means that for massive BHs ($M \gtrsim 10^4 M_{\odot}$) superradiance will be 
exponentially suppressed because the effective coupling $M\omega_p\gg1$~\cite{Brito:2015oca}. While it may be possible for environmental factors (e.g. due to dynamical or geometrical 
effects~\cite{Cardoso:2020nst,Cannizzaro:2020uap}) to suppress the contribution to the effective photon mass, even if superradiance could be triggered, Compton scattering will be efficient in quenching superradiance long before extracting a significant fraction of energy~\cite{Blas:2020kaa} (notice that the results presented in Fig.~\ref{fig:kin_mix} for the dark photon can be applied to the SM photon with a bare mass in the limit  $n_e \rightarrow 0$ and $\sin\chi_0 \rightarrow 1$); as such, superradiance cannot be used to probe the bare photon mass, should it exist.

\section{Conclusion}\label{sec:con}

In this work we have investigated the validity of superradiant bounds on dark photons, and argued there exists large regions of parameter space in which the dark photon interactions quench the superradiant growth before any observable change to the BH spin has occurred; specifically, we have shown that this occurs for vacuum kinetic mixings typically on the order of $\chi_0 \gtrsim 10^{-8}$, however this value depends on the mass of dark photon, the mass of the BH, and the local electron number density.  Additionally, we comment on other model-dependent quenching mechanisms, such as the minimum fermion mass required for a quenching via the Schwinger mechanism, or the restoration of the gauge symmetry which could arise in  models where the dark photon arises from the Higgs mechanism. While such mechanisms are interesting in their own right, they require additional model assumptions beyond the simple assumption of kinetic mixing made here.

While dark photons undergoing premature quenching of the superradiant growth will likely leave no discernible imprint on the BH spin distribution, we find that such particles may be capable of generating enormous bursts of light, with time-averaged luminosities extending up to $\sim 10^{57} \,{\rm erg / s}$. This signal should appear as a time- and energy-oscillating line, which could provide a striking signature of this phenomenon. This provides an alternative observational strategy for identifying ultra light bosons using BH superradiance.

The role of particle interactions in superradiant growth offers a rich and largely unexplored phenomenology. This work 
has focused on a single well-motivated extension of the SM, and showed that large regions of parameter space 
naturally evade conventional superradiant constraints, opening up new opportunities for detection. We hope that this 
will serve as a future guide toward understanding novel signatures of BH superradiance.

\subsubsection*{Acknowledgments}
We thank Aditya Parikh, Georg Raffelt, Toni Riotto, and Giuseppe Rossi for useful discussions. We also thank Edoardo Vitagliano and Enrico Cannizzaro for discussions and comments on the manuscript.
AC acknowledges support from the Israel Science Foundation (Grant No. 1302/19), the US-Israeli BSF (grant 2018236) and 
the German Israeli GIF (grant I-2524-303.7). A.C acknowledges also hospitality and support from the MPP of Munich. SJW 
acknowledges funding from the European Research Council (ERC) under the European Union's Horizon 2020 research and 
innovation programme (Grant agreement No. 864035 - UnDark).
PP acknowledges financial support provided under the European Union's H2020 ERC, Starting 
Grant agreement no.~DarkGRA--757480. We also acknowledge support under the MIUR PRIN and FARE programmes (GW-NEXT, 
CUP:~B84I20000100001), and from the Amaldi Research Center funded by the MIUR program ``Dipartimento di 
Eccellenza'' (CUP: B81I18001170001).

\appendix

\section{Plasma frequency around BHs} \label{app:plasma}

In this appendix we estimate the typical plasma density expected in the surrounding of an astrophysical BH. 
We use a simple Bondi-Hoyle model for spherical accretion onto a BH moving relative to a background baryonic density,
\begin{equation}
	\frac{d M}{dt} = {4\pi \lambda \rho_\infty}{v_{\rm eff}} r_B^2\, ,
\end{equation}
where $\lambda={\cal O}(1)$ is a model-dependent factor that accounts for the non-gravitational suppression of accretion, $\rho_\infty$ is the gas density far from the BH, $v_{\rm eff} = \sqrt{c_\infty^2 + v_{\rm rel}^2}$, where $c_\infty$ and $v_{\rm rel}$ are the speed of sound and the relative BH-baryon velocity far from the BH, respectively, and
\begin{equation}
 r_B = \frac{M}{v_{\rm eff}^2}
\end{equation}
is the Bondi radius. For typical non-relativistic effective velocities, the Bondi radius is much bigger than the BH gravitational radius, i.e. $r_B\gg M$.
In this regime the total infalling flow of baryons towards the BH is roughly constant, i.e.
\begin{equation}
n_e(r) v_{\rm ff}(r)r^2 = {\rm const} \,,
\end{equation}
where $v_{\rm ff}(r)=\sqrt{v^2_{\rm eff}+\frac{2M}{r}-\frac{2M}{r_B}}$ is the free fall velocity of the gas, which reduces to $v_{\rm eff}$ at the Bondi radius. Therefore, the electron density profile within the Bondi sphere reads
\begin{equation}
 n_e(r) = n_e^{\infty} \frac{v_{\rm eff}r_B^2}{v_{\rm ff}(r) r^2}\,,\label{densityprof}
\end{equation}
where $n_e^{\infty}=\rho_\infty/m_e$.
At distances $r\sim {\cal O}(M)$, $v_{\rm ff}(r)\sim \sqrt{2M/r}\gg v_{\rm eff}$, so that 
\begin{equation}
 n_e(r) \sim n_e^{\infty} \frac{v_{\rm eff}}{\sqrt{2M}} \frac{r_B^2}{ r^{3/2}}\,,\label{densityprof2}
\end{equation}
Finally, near the peak of the BH-boson condensate, $r_{\rm cloud}\sim M/\alpha^2$, we get 
\begin{equation}
 n_e(r_{\rm cloud}) \sim \frac{n_e^{\infty} }{\sqrt{2}v_{\rm eff}^3} \alpha^3 \approx 7\times10^2 n_e^{\infty} \left(\frac{\alpha}{0.1}\right)^3 \left(\frac{0.01}{v_{\rm eff}}\right)^3\,,\label{densityprofF}
\end{equation}
which is valid provided $r_{\rm cloud}\lesssim r_B$, i.e. $\alpha\gtrsim v_{\rm eff}$. The above estimate shows that the typical electron density in the relevant region near the BH can be several orders of magnitude larger than the ``ambient'' density at infinity, $n_e^{\infty}$, depending on $\alpha$ and $v_{\rm eff}$. Correspondingly, the plasma density $\omega_p\propto n_e^{1/2}$ can also vary by some orders of magnitude, see Eq.~\eqref{omegap}.
In the main text we keep the local density $n_e$ near the BH as a free parameter in the range $n_e\in(10^{-4},10^4){\rm cm}^{-3}$, which should bracket the model uncertainties.

\section{Non-linear regime}\label{app:KawDawson}
In this appendix we re-derive the results of Refs.~\cite{doi:10.1063/1.1692942, Akhiezer1956THEORYOW}, extending them 
to the electrodynamics with a massive dark photon. We can write Maxwell equations for both SM photon:
\begin{eqnarray}
\nabla \times \vec{E} = -\partial\vec{B}/\partial t, \\
\nabla \cdot \vec{B} = 0, \\
\nabla \times \vec{B} = 4\pi n_e e \vec{v}_e + -\partial\vec{E}/\partial t ,\\
\nabla \cdot \vec{E} = 4\pi e n_e,
\end{eqnarray}
and for a dark electromagnetic field:
\begin{eqnarray}
\nabla \times \vec{E'} = -\partial\vec{B'}/\partial t, \\
\nabla \cdot \vec{B'} = 0, \\
\nabla \times \vec{B'} = 4\pi n_e e \vec{v}_e + -\partial\vec{E'}/\partial t  - m_{\gamma'}^2 \vec{A}', \\
\nabla \cdot \vec{E} = 4\pi e n_e - m_{\gamma'}^2 A'^0.
\end{eqnarray}
These equations should be considered together with the electron equations of motion Eq.~\eqref{eq:electrons}, ignoring 
the collision term which we have seen to be not important for our goals. We want to study the generic problem of wave 
motions of the plasma, we therefore define the variable $\xi = \vec{i}\cdot \vec{r} - U t$, where $\vec{i}$ is a 
constant unit vector and $U$ the phase velocity of the wave. Using this new variable the equation of motion for the 
electrons becomes
\begin{eqnarray}
(\vec{i}\cdot \vec{v}_e - U )\partial \vec{p}_e/\partial_{\xi} = e  \vec{E}_{\rm obs} +e   \vec{u}_e \times \vec{B}_{\rm obs},
\end{eqnarray}
where we defined the observable fields $\vec{B}_{\rm obs} = \vec{B}+ \sin\chi_0 \vec{B'} $, 
$\, \vec{E}_{\rm obs} = \vec{E}+ \sin\chi_0 \vec{E'}$.

Now take the vector product of this equation with the wave vector $\vec{i}$ and consider only transverse waves ($\vec{i}\cdot \vec{B}_{\rm obs} = 0$) to get
\begin{equation}
(\vec{i}\cdot \vec{v}_e - U )\vec{i} \times  \dot{\vec{p}}_e = e \, \vec{i} \times \vec{E}_{\rm obs} -e \,\, \vec{i}\cdot \vec{v}_e  \vec{B}_{\rm obs},
\label{eq:DerivativePrime}
\end{equation}
where the overdot indicates a derivative respect to the wave variable $\xi$.
Using Faraday's law, we can also derive
\begin{eqnarray}
\vec{i} \times \vec{E}_{\rm obs} = U \vec{B}_{\rm obs},
\label{eq:FromFaraday2}
\end{eqnarray}
where we neglected background magnetic fields, as we are interested in a plasma which performs self-oscillations. We 
then consider Amper's law, multiplied vectorially on the left by $\vec{i}$ to obtain
\begin{align}
\dot{\vec{B}}_{\rm obs} = \frac{4\pi (1+\sin\chi _0^2)e n_e}{(U ^2 - 1)} \vec{i} \times \vec{v}_e - 
\frac{\sin\chi_0m_{\gamma'}^2}{U^2 - 1}\vec{i}\times \vec{A}',
\end{align}
where the dark vector potential can be related to the dark magnetic field via $\vec{i} \times \dot{\vec{A}}' = 
\vec{B'}$. This latter term is important in the limit of $m_{\gamma'} \gg \omega_p$, when the propagating light mode 
state -- triggering superradiance -- is aligned with $A'$ and we can write $\vec{A}' \simeq - 1/(e \sin\chi_0) 
\vec{p}_e$. Consequently, we can take the derivative of Eq.~\eqref{eq:DerivativePrime} to arrive to a second order 
differential equation for the electron momentum. In order to directly compare with Refs.~\cite{doi:10.1063/1.1692942, 
Akhiezer1956THEORYOW} let us rescale our variables, defining $\vec{\rho} = \vec{p}_e /m_e $ and rescaling the wave 
variable $\xi \rightarrow \xi /\omega_p$. We then get
\begin{equation}
\frac{d^2\vec{\rho}}{d\xi^2} + \frac{\vec{\rho}}{(\beta^2-1)\sqrt{1 + \rho^2}}\Big(1 + \frac{m_{\gamma'}^2}{\omega_p^2}\sqrt{1 + \rho^2}\Big) = 0,
\end{equation}
which is our master equation. The first term in the brackets is the one relevant when $m_{\gamma'} \ll \omega_p$, while the second one should be considered only in the regime $m_{\gamma'} \gg \omega_p$.

 Without loss of generality we can limit to consider the simple case where $\vec{i} = \hat{z}$, obtaining the solutions 
\begin{eqnarray}
\rho_x = \rho \cos\Big(\omega(t - z/U)\Big),\\
\rho_y = \rho \sin\Big(\omega(t - z/U)\Big).
\label{eq:sol_rho}
\end{eqnarray}

The frequency of oscillations is different depending on the hierarchy between $m_{\gamma'}$ and $\omega_p$. When $\omega_p \gg m_{\gamma'}$, we find a dielectric function $\epsilon(\omega) \equiv 1/U^2$ given by 
\begin{equation}
\epsilon(\omega) \simeq 1 - \omega_p^2 (1+\rho^2)^{-1/2}/\omega^2.
\end{equation}
If instead $\omega_{pl} \ll m_{\gamma'}$ to start with, the shortest time scale in the problem is always dictated by the oscillation of the dark fields and the dielectric function then simply reads 
\begin{equation}
\epsilon(\omega) \simeq 1 - m_{\gamma'}^2/\omega^2.
\end{equation}
 
The modulus $\rho$ can be connected to the observed magnetic and electric field using $\vec{i}\times \dot{\vec{p}} = - e 
\vec{B}_{obs} $ and $\vec{B} = 1/U\, \vec{i}\times \vec{E}$. We then find 
\begin{equation}
 \rho^2 = e^2 E_{\rm obs}^2/m_e^2 \omega^2\,.
\end{equation} 
We notice now that by definition the velocity of the electrons is $u_e^i = \rho^i /\gamma$, where $\gamma$ is the boost 
factor. We can then plug this solution for the electron momentum back into the spatial current term of Proca equations 
Eq.~\eqref{eq:Proca_equations}. Working in the non relativistic limit, we find the mass matrix in 
Eq.~\eqref{eq:mixingEq_3}, where the boost factor is a time-averaged one, due to the sinusoidal functions in 
Eq.~\eqref{eq:sol_rho}. A large boost factor for the electrons lower the plasma frequency as in the SM case. When the 
electrons move very fast, it will be very hard to have and excite any collective mode.

\bibliography{Ref}
\end{document}